\documentstyle[12pt,aaspp]{article}
\def\lsim{\raise2.90pt\hbox{$\scriptstyle
<$} \hspace{-6pt}\lower.5pt\hbox{$\scriptscriptstyle\sim$}\; }

\begin{document}
\title{MEASURING POLARIZATION IN THE COSMIC MICROWAVE BACKGROUND}

\author{Uro\v s Seljak}
\affil{ Harvard Smithsonian Center For Astrophysics, Cambridge, MA 02138 USA}
\begin{abstract}
Polarization induced by cosmological scalar perturbations 
leads to a typical anisotropy pattern, 
which can best be analyzed in Fourier domain. This 
allows one to unambiguously 
distinguish cosmological signal of polarization from 
other foregrounds and systematics, as well as from polarization
induced by non-scalar perturbations.
The precision with which 
polarization and cross-correlation
power spectra can be determined is limited by cosmic variance,
noise and foreground residuals. 
Choice of estimator can significantly improve our 
capability of extracting cosmological signal and
in the noise dominated limit the optimal power spectrum 
estimator 
reduces the variance by a factor
of two compared to the simplest estimator. If foreground residuals
are important  
then a different estimator can be used, which 
eliminates systematic effects from foregrounds so that no further 
foreground subtraction is needed. 
A particular combination of Stokes $Q$ and $U$ parameters vanishes
for scalar induced polarization, thereby allowing an unambiguous 
determination of tensor modes.
Theoretical predictions of polarization in standard models  
show that one typically expects a signal at the
level of 5-10$\mu$K on small angular scales and around 1$\mu$K on 
large scales ($l<200$). Satellite missions
should be able to reach sensitivities needed for
an unambiguous detection of polarization, which would help to break the 
degeneracies in the determination of some of the cosmological parameters.
\end{abstract}

\newpage
\section{Introduction}
Anisotropies in cosmic microwave background (CMB) are now widely accepted
as the best probe of early universe, which can potentially provide
information over a whole range of cosmological parameters 
(Jungman et al. 1996; Zaldarriaga, Spergel \& Seljak 1996). The main
advantage of CMB anisotropies as opposed to other, more
local probes, is that they are sensitive to the universe in the linear
regime, which statistical properties
can easily be calculated starting from ab-initio
theoretical
models and compared to the observations. A dozen or so cosmological
parameters could be extracted from 
the observations produced by the future satellite
and ground based experiments. There are two potential 
problems in this program. First is the somewhat
uncertain amount of galactic and
extragalactic foregrounds, which 
could severely limit our ability to extract cosmological signal from the
data. Second is the degeneracies among some of the cosmological 
parameters, which allows only certain combinations  
to be determined accurately, but does not allow to break 
the degeneracies between them (Bond et al. 1995; Jungman et al. 1996; 
Zaldarriaga et al. 1996).

It is therefore important to investigate other independent
confirmations of results produced from CMB anisotropies and it has
long been recognized that polarization in the microwave sky might
provide such an independent test (Rees 1968, Polnarev 1985, 
Bond \& Efstathiou 1987,
Crittenden, Davis \& Steinhardt 1993, Frewin, Polnarev \& Coles 1994, 
Coulson, Crittenden \& Turok 1994, Crittenden, Coulson \& Turok 1995). 
Like temperature anisotropy,
polarization probes the universe in the linear regime and so can 
provide information useful to determine cosmological parameters.
It is specially
important for determining parameters which are only weakly constrained
by the CMB anisotropies alone, such as the epoch of reionization or
the presence of tensor perturbations. In both of these 
cosmic variance is the limiting factor in our capability of
extracting the parameters, so measuring polarization
would increase the amount of information and allow for
a more accurate determination of the parameters.  
Both polarization-polarization and polarization-temperature
correlation give an independent set of power
spectra, which have different sensitivity to different parameters, 
so combining them results in a much larger
information about the underlying 
cosmological model and can significantly increase our capability 
of extracting useful information from the CMB measurements. 

The main disadvantage of polarization is that 
it is predicted to be of a rather low
amplitude, of the order of 10\% of temperature 
anisotropy, so measuring it
represents an experimental challenge that has yet to be overcome.
Currently there are no positive detections of polarization, with the
best upper limits of the order of $25\mu K$ (Wollack et al. 1993).
This situation needs not continue in the future, however, as the
experimental sensitivity increases and new larger and
better experiments are being planned. Moreover, as shown in this 
paper, cosmologically induced 
polarization has a unique signature in the data that cannot be mimicked
by other foregrounds and provides a clear test of the searched 
signal. An
experiment at Brown (Timbie 1996) plans to reach
sensitivities of a few $\mu K$, which should be sufficient to detect the
polarization
signal if our current expectations are correct. Interferometer
observations could be able to extend this to much larger
areas of the sky and produce maps of polarization both in real space and
in frequency space. Unfortunately, at present none of the planned
interferometer 
experiments is including  polarization, although as argued in this paper 
frequency space has many
advantages in search for the unique signature of cosmological polarization.
Finally, all sky satellite maps of polarization are also planned
by MAP and COBRAS/SAMBA satellites and may eventually 
provide us with very high accuracy maps of polarization pattern in the sky. 

The outline of this paper is the following. In \S 2 statistical
properties of polarization parameters are derived in 
the small scale limit. Special nature of polarization induced by 
scalar perturbations allows one to use statistical methods developed
in the context of weak lensing (Kaiser 1992). In \S 3 various 2-point 
estimators are presented both in Fourier and
real space, together with their variances and covariances. 
\S 4 is devoted to the foregrounds and possible methods of their
elimination. Polarization induced by tensor modes is discussed
in \S 5 and the difference between the two types of perturbations
is highlighted. 
In \S 6 theoretical predictions for polarization
are computed for a variety of cosmological models and the ability to 
extract them with the satellite missions is discussed. This is
followed by conclusions in \S 7. 

\section{Polarization in the small scale limit}
In this section we derive the small-scale limit of temperature 
and polarization anisotropies.
This limit is of special interest, because
one can replace the general spherical expansion with
the Fourier expansion and the
expressions simplify considerably. 
The analysis in this section will be restricted to polarization
generated by scalar perturbations.   
Tensor perturbations are discussed in \S 5.

In general one can describe CMB polarization as
a 2x2 temperature perturbation tensor $T_{ij}$. Stokes parameters $Q$ and
$U$ (we will ignore $V$ in the following, since it cannot be generated through
Thomson scattering) are defined as $Q=T_{xx}-T_{yy}$ and
$U=2T_{xy}=2T_{yx}
$, while temperature anisotropy is just its trace, $T=T_{xx}+T_{yy}$.
The components 
are defined with respect to a fixed coordinate system $(x,y)$
perpendicular to the photon direction $\vec{n}$.
The Stokes parameter $Q$ is positive if temperature perturbation
is larger
along the $x$ relative to the $y$ axis,
while $U$ parameter is positive
if perturbation is larger along the upper right diagonal relative to the
upper left diagonal.

Equations of radiative transfer for 
polarization and temperature anisotropy 
simplify in Fourier space if we work in a 
frame with the 
axis defined parallel and perpendicular to $\partial/\partial \beta$,
where $\beta$ is the angle between the wavevector $\vec{k}$ and photon
direction $\vec{n}$. 
Azimuthal symmetry guarantees that the
Thomson scattering preserves the diagonal form of $I_{ij}$, just like in 
the case of plane-parallel atmospheres (Chandrasekhar 1960,
Kaiser 1983).
In terms of the Stokes parameters
only $Q$ is excited in this frame. This puts a restriction on the general
form of polarization and as we show below it can be used to separate
cosmologically induced polarization from other sources and systematic
effects. Although only $Q$ is present in this frame, when 
we rotate polarization by an azimuthal 
angle $-\phi_{\vec{k},\vec{n}}$ to a fixed frame in the sky
we generate both $Q$ and $U$.
At the observers position the expressions
for $T$, $Q$ and $U$ 
are given by (Bond \& Efstathiou 1987, Kosowsky 1996),
\begin{eqnarray}
T(\vec{n})&=&\int d^3\vec{k}\Delta_T(\vec{k},\vec{n})\nonumber \\
Q(\vec{n})&=&\int d^3\vec{k}\Delta_P(\vec{k},\vec{n})\cos(2\phi_{\vec{k},\vec{n}
})
\nonumber \\
U(\vec{n})&=&\int d^3\vec{k} \Delta_P(\vec{k},\vec{n})
\sin(2\phi_{\vec{k},\vec{n}}),
\label{tqu}
\end{eqnarray}
where $\Delta_{T,P}(\vec{k},\vec{n})$ are the Fourier components of
temperature and 
polarization distribution function integrated over the momentum 
and $\vec{n}$ is the direction of observation on the sky. 
The expression for rotation angle $\phi_{\vec{k},\vec{n}}$
depends both on $\vec{k}$ and $\vec{n}$, but if
we restrict our attention to the directions $\vec{n}$ around the pole then 
it can be approximated with $-\phi_k$, where $\phi_k$ is 
the azimuthal angle of vector $\vec{k}$.
This approximation breaks down for wavemodes $\vec{k}$ close to the pole
($\hat{z}$ direction), but for sufficiently small scales the contribution
from these modes to the total power becomes negligible.

The solution for $\Delta_{T,P}(\vec{k},\vec{n})$ can be written as an 
integral over the sources 
along the line of sight (Seljak \& Zaldarriaga 1996)
\begin{equation}
\Delta_{T,P}(\vec{k},\vec{n})
=\int_0^{\tau_0}d\tau e^{i \vec{k}\cdot \vec{n} (\tau -\tau_0)}
S_{T,P}(k,\tau), 
\label{formal}
\end{equation}
where $S_{T,P}(k,\tau)$ are the source functions for temperature and
polarization and can be expressed in terms of metric, baryon and 
photon perturbations (see Seljak \& Zaldarriaga 1996 for their 
explicit expressions). 
By combining equations \ref{tqu} and \ref{formal} we can write the 
complete solutions for $T(\vec{n})$, 
$Q(\vec{n})$ and $U(\vec{n})$. 
The only term that depends on the direction
$\vec{n}$ in equation \ref{formal} is the exponential.
The expressions for polarization can therefore be 
simplified by noting that for the directions $\vec{n}$ near the pole
$\cos(2\phi_{\vec{k}})$ and $\sin(2\phi_{\vec{k}})$ can be written
in terms of second derivatives with respect to $\vec{\theta}$,
the 2-dimensional 
projection of $\vec{n}$
onto fixed $(x,y)$ plane perpendicular to the pole. This leads to
\begin{eqnarray}
Q(\vec{n})&=&D_Q(\vec{n})\int d^3\vec{k}\Delta_P(\vec{k},\vec{n}),\,\,\,\,\,\,\,\,\,\,\,\,\,\,\,
D_Q(\vec{n})=(\partial \theta_x\partial \theta_x-
\partial \theta_y\partial \theta_y)\nabla_\theta^{-2}
\nonumber \\
U(\vec{n})&=&D_U(\vec{n})\int d^3\vec{k} \Delta_P(\vec{k},\vec{n}) ,\,\,\,\,\,
\,\,\,\,\,\,\,\,\,\,
D_U(\vec{n})=2\partial \theta_x\partial \theta_y\nabla_\theta^{-2},
\end{eqnarray}
where $\nabla_{\theta}^{-2}$ is the inverse 2-d Laplacian with respect 
to $\vec{\theta}$ and $\theta_x$, $\theta_y$ are the components of
$\vec{\theta}$ in the fixed basis on the sky. 
Because $\Delta_{T,Pl}(
\vec{k},\vec{n})$ depend only on the angle between the two vectors
one can 
expand them in Legendre series,
\begin{equation}
\Delta_{T,P}(\vec{k},\vec{n})=\sum_l(2l+1)(-i)^l\Delta_{T,Pl}(k)P_l(
\mu),
\end{equation}
where $\mu=\vec{k}\cdot \vec{n}$.
The rms values for $\Delta_{T,Pl}(k)$ can be obtained
by solving the Boltzmann equation in differential form or the integral solution
itself (Bond \& Efstathiou 1987, Seljak 
\& Zaldarriaga 1996).

Each of the observable quantities $T(\vec{n})$, 
$Q(\vec{n})$ and $U(\vec{n})$
can be expanded on a sphere into spherical 
harmonics or their derivatives,
\begin{eqnarray}
T(\vec{n})&=&\sum_{lm}a_{Tlm}Y_{lm}(\vec{n}) \nonumber \\
Q(\vec{n})&=&\sum_{lm}a_{Plm}D_Q(\vec{n})Y_{lm}(\vec{n}) \nonumber \\
U(\vec{n})&=&\sum_{lm}a_{Plm}D_U(\vec{n})Y_{lm}(\vec{n}) \nonumber \\
a_{T,Plm}&=&4\pi (-i)^l \int d^3\vec{k} Y^*_{lm}(\vec{k})\Delta_{T,Pl}(\vec{k}).
\label{alm}
\end{eqnarray}
The statistical properties of $a_{T,Plm}$ coefficients follow from equation 
\ref{alm} above,
\begin{eqnarray}
\langle a_{Tlm}^* a_{Tl'm'} \rangle &=&
\delta_{ll'} \delta_{mm'} 
\int d^3k \Delta^2_{Tl}(k) 
\equiv \delta_{ll'} \delta_{mm'} C_{Tl}\nonumber \\
\langle a_{Plm}^* a_{Pl'm'} \rangle &=&
\delta_{ll'} \delta_{mm'} 
\int d^3k \Delta^2_{Pl}(k) 
\equiv \delta_{ll'} \delta_{mm'} C_{Pl}
\end{eqnarray}
The cross-correlation between temperature and polarization is given by
\begin{equation}
\langle a_{Tlm}^* a_{Pl'm'} \rangle=\delta_{ll'} \delta_{mm'}
\int d^3k \Delta_{Tl}(k)\Delta_{Pl}(k)
\equiv \delta_{ll'} \delta_{mm'}C_{Cl}.
\label{cl}
\end{equation}

Because we are only interested in $\vec{n}$ near the pole one can
approximate the sphere locally as a plane, in which case
instead of spherical decomposition we may use plane wave expansion. 
In this limit we replace $\sum_{lm}a_{Plm}Y_{lm}(\vec{n})$ with 
$\int d^2\vec{l} P(\vec{l})e^{i\vec{l}\cdot\vec{\theta}}$ (and 
analogously for temperature anisotropy\footnote{A somewhat 
better correspondence between small scale and large 
scale expressions is achieved if one uses $l+1/2$ as the amplitude 
of a wavevector that corresponds to $C_l$ (Bond 1996).}). Differential 
operators $D_Q(\vec{n})$ and $D_U(\vec{n})$ acting on $e^{i\vec{l}\cdot\vec{\theta}}$ become simple again 
and bring out 
$\cos (2\phi_{\vec{l}})$ and $\sin(2\phi_{\vec{l}})$, respectively, 
where $\phi_{\vec{l}}$ is the direction
angle of 2-dimensional vector $\vec{l}$ with amplitude $l$. 
This leads to
\begin{eqnarray}
T(\vec{\theta})&=&(2\pi)^{-2}\int d^2\vec{l}e^{i\vec{l}\cdot \vec{\theta}
}T(\vec{l}) \nonumber \\
Q(\vec{\theta})&=&(2\pi)^{-2}\int d^2\vec{l}e^{i\vec{l}\cdot \vec{\theta}
}P(\vec{l})\cos(2\phi_{\vec{l}}) 
\nonumber \\
U(\vec{\theta})&=&(2\pi)^{-2}\int d^2\vec{l}e^{i\vec{l}\cdot \vec{\theta}}
P(\vec{l})\sin(2\phi_{\vec{l}}). 
\end{eqnarray}
$T(\vec{l})$ and $P(\vec{l})$ are the Fourier components of temperature 
anisotropy and polarization in $\vec{l}$ space and have
the statistical properties, 
\begin{eqnarray}
\langle T(\vec{l})T^*(\vec{l'}) \rangle =(2\pi)^2C_{Tl}\delta_D(l-l') 
\nonumber \\
\langle P(\vec{l})P^*(\vec{l'}) \rangle =(2\pi)^2C_{Pl}\delta_D(l-l') 
\nonumber \\
\langle T(\vec{l})P^*(\vec{l'}) \rangle =(2\pi)^2C_{Cl}\delta_D(l-l'), 
\end{eqnarray}
where $\delta_D(l-l')$ is 
the Dirac $\delta$ function as opposed to the Kronecker $\delta$ 
in the discrete case and $C_l$ is assumed to be a continuous function
obtained by interpolation from the discrete spectrum defined in 
equation \ref{cl}.

To generate a map of temperature anisotropy and polarization one proceeds
in the following way. For each pair of vectors $\vec{l}$, $-\vec{l}$
on a discrete mesh one
diagonalizes the correlation matrix $M_{11}=C_{Tl}$, 
$M_{22}=C_{Pl}$, $M_{12}=M_{21}=C_{Cl}$, where $l$ is the 
amplitude of vector $\vec{l}$. One
then generates from a normalized gaussian distribution
two pairs of 
random numbers and multiplies them with the amplitudes given by the 
square root of 
the correlation matrix eigenvalues. Rotating this vector pair back to
the original frame gives a realization of $T(\vec{l})$ and $P(\vec{l})$ 
(and their complex conjugates corresponding to $-\vec{l}$), 
from which follow
$Q(\vec{l})=P(\vec{l})\cos(2\phi_{\vec{l}})$ 
and $U(\vec{l})=P(\vec{l})\sin(2\phi_{\vec{l}})$. Fourier transform of 
$T$, $Q$ and $U$ back into the real space 
gives a map of these quantities in the sky in the 
small-scale limit, with the correct auto and cross 
correlations among all the quantities. 
Note that this differs from the prescription given 
in Bond \& Efstathiou 1987.

\section{2-point estimators}

The 2-point correlations can be calculated either in angular or in 
frequency (Fourier) 
space. While the two are related via a Fourier transform, 
there are certain advantages to the analysis performed in frequency 
space. In the first subsection we explore this approach, while the correlation 
function approach and the comparison between the two 
are explored in the next subsection.  

\subsection{Power spectrum analysis}

From a map of $T(\vec{\theta})$, $Q(\vec{\theta})$ and $U(\vec{\theta})$ 
we can obtain 
their analogs in frequency space by 
Fourier transform,
$X(\vec{l})=\int d^2\vec{\theta} e^{-i\vec{l}\cdot \vec{\theta}}X(\vec{\theta})$,
where $X$ stands for $T$, $Q$ or $U$. 
Using the expressions given in the previous section one obtains the
two-point functions of these quantities,
\begin{eqnarray}
\langle T(\vec{l}) T^*(\vec{l'})\rangle &=&(2\pi)^2C_{Tl}\delta_D(l-l') \nonumber \\
\langle T(\vec{l}) Q^*(\vec{l'})\rangle&=&(2\pi)^2C_{Cl}\cos(2\phi_{\vec{l}})
\delta_D(l-l') \nonumber \\
\langle T(\vec{l}) U^*(\vec{l'})\rangle&=&(2\pi)^2C_{Cl}\sin(2\phi_{\vec{l}})
\delta_D(l-l') \nonumber \\
\langle Q(\vec{l}) Q^*(\vec{l'})\rangle&=&(2\pi)^2C_{Pl}\cos^2(2\phi_{\vec{l}})
\delta_D(l-l') \nonumber \\
\langle U(\vec{l}) U^*(\vec{l'})\rangle&=&(2\pi)^2C_{Pl}\sin^2(2\phi_{\vec{l}})
\delta_D(l-l').
\end{eqnarray}
We see that the correlations in polarization give rise to a very 
characteristic anisotropy pattern in $\vec{l}$ space. This arises from 
the fact that polarization was not generated from a general mechanism, 
but rather through a process of 
Thomson scattering, which cannot generate $U$ and 
$V$ components in the $\vec{k}$ dependent frame defined in previous
section. This characteristic anisotropy can therefore be used to separate true 
signal from instrumental artifacts and foregrounds, which is discussed
in more detail in next section. 
Alternatively, 
if foregrounds can be kept
under control one can use the characteristic anisotropy to separate scalar
induced polarization from the one induced by vector or 
tensor modes, which do not
obey the same anisotropy pattern (\S 5).

To estimate the sensitivity that is possible to achieve in a measurement
of polarization power spectrum let us assume the
measurements are given on a square grid of $N$
pixels with a total solid angle $\Omega$. The antenna 
beam smearing will be described 
with $b(l)$. In the case of single dish observations with 
a gaussian beam this
is given by $b(l)=e^{l^2\sigma_b^2/2}$, where $\sigma_b$ is the gaussian size of
the beam. In the case of interferometers $b(l)$ is either 1 or 0, 
depending on whether the particular frequency is observed or not.
We will assume that different measurements are uncorrelated by taking 
mesh spacing large enough to ignore correlations induced by finite
window (Hobson \& Magueijo 1996). 
In the case of single dish experiments each pixel in real space has a noise 
contribution  
with rms noise amplitudes 
$\sigma_T$, $\sigma_P$ for 
temperature and both components of 
polarization, respectively (we assume for simplicity 
that $Q$ and $U$ are being measured equal amounts of time). 
We will also assume that noise is uncorrelated between different pixels
and between different polarization components
$Q$ and $U$. This is only 
the simplest possible choice and more complicated noise correlations
arise if all the components are obtained from a single set of 
observations. 
In the case of Brown polarization experiment (Timbie 1996) the
polarization measurements will be made by rotating the antenna axis by 
$45^\circ$, each time measuring directly the difference between the two 
orthogonal polarizations. This would therefore provide a direct measurement 
of $Q$ and $U$ components with no noise 
mixing between them. In the case of interferometers each pixel in
$\vec{l}$ space is measured directly and the noise is uncorrelated 
between individual 
pixels in frequency space. 
Following Knox 1995 we will introduce pixel independent
measure of noise $w^{-1}_{T,P}=\Omega \sigma^2_{T,P}/N$ for single dish
experiments and $w^{-1}_{T,P}= \sigma^2_{T,P}$ for interferometers 
(see Hobson \& Magueijo 1996 for expressions that relate $\sigma_{T,P}$
to the receiver sensitivity).
For cross-correlation between temperature and polarization there 
are two simple cases to be considered. In one the cross-correlation
is being made with two different maps, in which case noise is 
uncorrelated. In the second case both temperature and polarization are
obtained from the same experiment by adding and differenciating 
the two polarization states.
In this case noise in temperature and polarization are
related via $\sigma_T^2=\sigma_P^2/2$. 
In both cases noise in temperature is uncorrelated with noise in
polarization components, so the final expressions are identical.
 
The first step is to construct a discrete Fourier transform of the map,
$\hat{X}(\vec{l})=
N^{-1}\sum e^{-i\vec{l}\cdot \vec{\theta}}X(\vec{\theta})$.
Using the observed quantities $\hat{T}(\vec{l})$, $\hat{Q}(\vec{l})$ and 
$\hat{U}(\vec{l})$ 
we can form power spectrum estimates. The simplest one is for 
temperature anisotropy, which for single dish observations is given by
\begin{equation}
\hat{C}_{Tl}=\left[\sum_l {\Omega \over N_l}[\hat{T}(\vec{l})\hat{T}^*(\vec{l})]
-w^{-1}_T \right]b^{-2}(l),
\end{equation}
where the term $b^{-2}(l)$ accounts for beam smearing and $N_l$ is the
number of independent modes around $l$. The expression for
interferometers is the same without the beam smearing term and summing 
only over the observed modes in frequency space. We will 
only present expressions for single dish here, as the modification for
interferometers is obvious. Each mode 
$\hat{T}(\vec{l})$ is
a complex random variable with 0 mean and variance 
$C_{Tl}b^{2}(l)+w^{-1}_T$. The variance on estimator $\hat{C}_{Tl}$ for 
a single pair of modes $\hat{T}(\vec{l})$, $\hat{T}(-\vec{l})$
is therefore $C_{Tl}b^{2}(l)+w^{-1}_T$. 
If we average over $N_l$ modes the variance is reduced by $N_l^{-1/2}$.
There are $l^2\Omega d\ln l/2\pi$ modes of amplitude $l$ in an interval 
$d\ln l$, which leads to the variance 
\begin{equation}
{\rm Cov }(C_{Tl}^2)=C_{Tl}^2
{4\pi \over l^2\Omega d\ln l}\left(1+[w_TC_{Tl}b^{2}(l)]^{-1}\right)^2,
\end{equation}
where ${\rm Cov}(XX') \equiv\langle (\hat{X}-\langle \hat{X} \rangle 
)(\hat{X}'-\langle \hat{X}'\rangle)\rangle$.
If there is 
more than one field the variance decreases inversely proportional to the
square root of number of fields
(Hobson \& Magueijo 1996). In the limit of large sky coverage
it reduces to the expressions
given by Knox 1995 and Jungman et al. 1996. 

In the case of polarization there are several estimators that one can 
form. The simplest one is given by 
\begin{equation}
\hat{C}_{Pl}=\left[ {\Omega \over N_l}\sum_l[\hat{Q}(\vec{l})\hat{Q}^*(\vec{l})
+\hat{U}(\vec{l})\hat{U}^*(\vec{l})]-2w^{-1}_P
\right] b^{-2}(l),
\label{est}
\end{equation}
The single mode pair 
variance for this estimator is $(C_{Pl}^2b^{4}(l)+2C_{Pl}b^{2}(l)w^{-1}_P
+4w^{-2}_P)^{1/2}$. If noise is dominant contributor to the variance, as it is 
likely to be the case for polarization given the small overall amplitude
of the signal, then this estimator is far from optimal. 
The optimal estimator is
\begin{equation}
\hat{C}_{Pl}=\left[{\Omega \over N_l}\sum_l |\hat{Q}(\vec{l})\cos(2\phi_{\vec{l}})
+\hat{U}(\vec{l})\sin(2\phi_{\vec{l}})|^2-w^{-1}_P
\right] b^{-2}(l),
\label{optimal}
\end{equation}
which has single mode pair 
variance $C_lb^{2}(l)+w^{-1}_P$. In the limit of 
noise dominated variance this is 2 times
smaller than the variance of the estimator in equation \ref{est}.  
Averaging over all modes in $d\ln l$ gives 
\begin{equation}
{\rm Cov}(C_{Pl}^2)=C^2_{Pl}
{4\pi \over l^2\Omega d\ln l}\left(1+[w_PC_{Pl}b^{2}(l)]^{-1}\right)^2.
\label{covpl}
\end{equation}

Finally, for cross-correlation the optimal estimator is,
\begin{equation}
\hat{C}_{Cl}={\Omega \over 2N_l}\sum_l\left[
\left( \hat{Q}(\vec{l})\hat{T}^*(\vec{l})
+\hat{Q}^*(\vec{l})\hat{T}(\vec{l})\right)\cos(2\phi_{\vec{l}})
+\left( \hat{U}(\vec{l})\hat{T}^*(\vec{l})
+\hat{U}^*(\vec{l})\hat{T}(\vec{l})\right)\sin(2\phi_{\vec{l}})\right]
b^{-2}(l),
\end{equation}
which has a single mode pair variance 
$[(C^2_{Cl}b^{4}(l)+(C_{Tl}b^{2}(l)+w_T^{-1})(C_{Pl}b^{2}(l)+
w_P^{-1}))/2]^{1/2}$.  
As before, averaging
over the modes reduces this variance inversely proportional to the square 
root of the number of modes and gives,
\begin{equation}
{\rm Cov}(C_{Cl}^2)=C^2_{Cl}
{2\pi \over l^2\Omega d\ln l}\left[
1+{(C_{Tl}b^{2}(l)+w_T^{-1})(C_{Pl}b^{2}(l)+w_P^{-1}) \over 
C^2_{Cl}b^{4}(l)}\right].
\label{covcl}
\end{equation}

For a study of cosmological parameters one also needs to include the
covariance elements between various power spectrum estimators. 
These are given by
\begin{eqnarray}
{\rm Cov}(C_{Tl}C_{Pl})&=&C_{Cl}^2(l){4\pi \over l^2\Omega d\ln l} 
\nonumber \\
{\rm Cov}(C_{Cl}C_{Tl})&=&C_{Cl}C_{Tl}{4\pi \over l^2\Omega d\ln l}
(1+[w_TC_{Tl}b^2(l)]^{-1}) \nonumber \\
{\rm Cov}(C_{Cl}C_{Pl})&=&C_{Cl}C_{Pl}{4\pi \over l^2\Omega d\ln l}
(1+[w_PC_{Pl}b^2(l)]^{-1}).
\end{eqnarray}
\subsection{Correlation function analysis}

In this subsection we explore the correlation function analysis of
CMB anisotropies. 
Taking Fourier transform of the power spectra leads to the following 
correlation functions,
\begin{eqnarray}
\langle T(0)T(\vec{\theta})\rangle &=&
\int {l dl \over 2\pi} C_{Tl}b^{2}(l) J_0(l\theta)\nonumber \\
\langle Q(0)Q(\vec{\theta})\rangle &=&
\int {l dl \over 4\pi} C_{Pl}b^{2}(l) [J_0(l\theta)+\cos(4\phi)J_4(l\theta)]\nonumber \\
\langle U(0)U(\vec{\theta})\rangle &=&
\int {l dl \over 4\pi} C_{Pl}b^{2}(l) [J_0(l\theta)-\cos(4\phi)J_4(l\theta)]\nonumber \\
\langle Q(0)U(\vec{\theta})\rangle &=&
\int {l dl \over 4\pi} C_{Pl}b^{2}(l)\sin(4\phi)J_4(l\theta)\nonumber \\
\langle T(0)Q(\vec{\theta})\rangle &=&
\int {l dl \over 4\pi} C_{Cl}b^{2}(l)\cos(2\phi)J_2(l\theta)\nonumber \\
\langle T(0)U(\vec{\theta})\rangle &=&
\int {l dl \over 4\pi} C_{Cl}b^{2}(l)\sin(2\phi)J_2(l\theta),
\label{corrf}
\end{eqnarray}
where $\phi$ is the direction angle of $\vec{\theta}$ and
$J_n(x)$ are the Bessel functions of order $n$.
Although the characteristic anisotropy is present also in the
correlation functions, it is more complicated, because there are
actually two independent correlation functions for polarization,
\begin{eqnarray}
C^1_{P}(\theta)&=&\int {l dl \over 4\pi} C_{Pl}b^{2}(l)J_0(l\theta) \nonumber \\
C^2_{P}(\theta)&=&\int {l dl \over 4\pi} C_{Pl}b^{2}(l)J_4(l\theta),
\end{eqnarray}
both of which of course depend on the same underlying power spectrum.
Moreover, 
the prior $C_{Pl}>0$ becomes a set of integral constraints in real space,
which is more difficult to impose on the estimators obtained from 
the data. All this argues for Fourier space analysis as the method of
choice in the case of polarization.

Expressions above agree with those derived
by Coulson et al. 1994 and Kosowsky 1996, which for $QQ$ and $UU$ 
are also only valid 
in the small scale limit, but where this limit has not been
consistently applied to all the steps, so that their final expressions
look much more complicated than they actually are.   
They contain a term
involving double summation over $\Delta_{Pl}$,
$\Delta_{Pl'}$ in the $\phi$ dependent term, which as shown in Appendix
reduces to the expressions above 
if one consistently applies small scale limit to their expression.
All the information about polarization is therefore
contained in $C_{Pl}$ and $C_{Cl}$. 
Note that taking the Fourier transform of $QQ$ or $UU$ correlation 
function does not result in the power spectrum of $C_{Pl}$, but has 
an additional $\phi$ dependent term that involves a double 
integral of $J_0(l\theta)$ and $J_4(l'\theta)$ 
over $l'$ and $\theta$. Although generally smaller than $C_{Pl}$,
this integral does not vanish in general, 
hence the appearance of such terms in expressions by Coulson et al.
(1994) and Kosowsky (1996). This is of course not the optimal 
way to obtain the power spectrum from the correlation function. 
To obtain an estimate of 
the underlying power spectrum 
it is better to  
work in the Fourier domain directly, 
following the methods given in previous subsection. 

\subsection{Predicting polarization from temperature maps}
One can use the measured temperature maps to predict the polarization 
pattern. The estimator is
\begin{equation}
\hat{Q}(\vec{l})=\alpha T(\vec{l})\cos(2\phi_{\vec{l}}), \, \, 
\, \, \hat{U}(\vec{l})=\alpha T(\vec{l})\sin(2\phi_{\vec{l}}),
\end{equation}
where by minimizing the variance 
\begin{equation}
\langle [\hat{Q}(\vec{l})-Q(\vec{l})]^2\rangle= \cos^2(2\phi_{\vec{l}})
[\alpha^2C_{Tl}-2\alpha C_{Cl}+C_{Pl}]
\end{equation}
and similarly for $U$ one finds 
$\alpha=C_{Cl}/C_{Tl}$, so that the fractional variance in the estimator 
is 
\begin{equation}
{\langle [\hat{Q}(\vec{l})-Q(\vec{l})]^2\rangle \over Q(\vec{l})^2}
=1-\rm{Corr}(T,P)_l^2.
\end{equation}
The correlation coefficient is defined as $\rm{Corr}(T,P)_l=C_{Cl}/(
C_{Tl}C_{Pl})^{1/2}$. 
Figure \ref{fig1}d shows that the correlation coefficient typically 
ranges between $-0.5$ and $0.5$ and so the fractional variance 
will be at best around 0.8 or so in $\vec{l}$ space and even larger than 
that in real space, where one averages over positive and negative 
cross-correlations in power spectrum. This is 
not very impressive in terms of predicting where to look for 
large polarization amplitude, although in a statistical sense 
one is still much more likely to find a high signal at a high 
$\sigma$ peak in the temperature map than at a random 
point (Coulson et al. 1994). On large angular scales ($l<10$)
the correlation 
coefficient can be much larger and approaches unity in some models, 
so that the expressions in this limit would
actually give a good correspondence between observed and predicted
polarization. Unfortunately, the amplitude of polarization is extremely
small on these scales and there is little hope to measure it in the
near future even with the help of this ``matching filter'' technique.

\section{Removal of foregrounds and other systematics}
Although several galactic and extragalactic 
foregrounds are significant in the case of
temperature measurements, only few of these are polarized
and need to be considered for polarization measurements. 
Radiation from earth atmosphere and  bremsstrahlung emission 
are not polarized at millimeter wavelengths
and need not be discussed further (although atmospheric 
emission does have an effect by increasing the effective 
temperature of receiver and adding a fluctuating offset). 

Extragalactic radio sources have synchrotron radiation as the
dominant emission mechanism and can be 20\% polarized.
Their contribution to the polarization signal will be similar
as their contribution to the temperature anisotropy signal and
so analysis of point source effects on CMB can be directly 
applied to the case of polarization as well. 
This is discussed in more detail by Francheschini et al. 
(1989, 1991) and Tegmark \& Efstathiou (1996). 
Point source contamination depends on the 
observed frequency, angular scale and flux cut above which 
point sources can be identified and eliminated. Poisson 
distribution produces a white-noise spectrum and at 
large angular scales radio point 
sources do not pose a significant problem.  
For example, on angular scales
above $1^\circ$ their contribution is less than $1\mu K$ at 30 GHz
(Timbie 1996), which is below the expected amplitude of the signal
and is even less than that at higher frequencies. 
On smaller angular scales point sources become more important and
more ambitious flux cuts are needed, which limits the area of the
sky that can be observed. While in the case of temperature anisotropy
this can be the main limitation of an experiment (and indeed of the
whole CMB field), in the case of polarization we do have an 
additional constraint that allows us to separate cosmologically 
induced signal from the foregrounds. This is discussed in more detail
below.

On large angular scales the main foreground contribution comes from our galaxy,
where both dust and synchrotron emission can be polarized.
Dust emission in the far-infrared is polarized 
up to 10\% (Hildebrand
et al. 1995).
The contribution of dust to the lower frequency channels (where most 
HEMT based polarization measurements are being planned) is small.
At frequencies below 100 Ghz the most important source of polarization
is likely to be galactic synchrotron emission.
Its linear polarization can
reach 70\%, so that polarization 
amplitudes of 50$\mu$K are expected around
30GHz (Cortiglioni \& Spoelstra 1995). This drops significantly at 
higher frequencies and only
a few $\mu$K synchrotron contamination in polarization is expected 
around 100GHz (Timbie 1996). Nevertheless, this is of the same order 
as the expected signal, so that some further foreground rejection is
needed. One possibility is to use only clean parts of the sky where 
synchrotron emission is low, such as at high galactic latitudes.
In addition,
multifrequency CMB observations can be used to remove the
foregrounds, which in the case of only one important foreground 
with approximately known frequency dependence can be very 
effective (Brandt et al. 1994). Accuracy of 1$\mu$K can be 
achieved with only two frequency channels if the noise
level is around 1$\mu$K per pixel (Timbie 1996). Another possibility is
multifrequency removal using the more sensitive temperature maps. 
This would be a useful strategy for example in the case of COBRAS/SAMBA 
satellite, where only lower frequency channels will have polarization
capabilities, but all frequency channels could be used to determine
the local contribution of various foregrounds. This way one
could effectively
remove multicomponent foregrounds from polarization 
even if only a few channels actually measured it.

While each of the foregrounds above can in principle be removed
from the data, in practice this may not always be possible at 
the levels of 1$\mu$K and it would be useful to have an 
additional test of the presence of cosmological signal. 
Characteristic anisotropy of polarization in Fourier space 
provides such a test and gives a unique signature of polarization
induced by scalar perturbations. 
To test whether the signal is cosmological one needs to compare the 
quantities 
\begin{equation}
E(\vec{l})=Q(\vec{l})
\cos(2\phi_{\vec{l}})
+U(\vec{l})\sin(2\phi_{\vec{l}})
\label{fore}
\end{equation}
and
\begin{equation}
B(\vec{l})=
-Q(\vec{l})
\sin(2\phi_{\vec{l}})
+U(\vec{l})\cos(2\phi_{\vec{l}}).
\end{equation}
The first quantity 
contains all the polarization signal and its estimator is given 
in equation \ref{optimal}, while the second
quantity vanishes even in the presence of cosmological polarization
induced by scalar perturbations. This is true not just statistically,
but for each Fourier mode individually. Most of the foregrounds should
contribute on average the same amount to both
variables. This is certainly valid for uncorrelated point 
sources, but even in the case of synchrotron and dust emission the alignment is 
preferentially determined by magnetic fields, which are not scalar in nature
and will not exhibit the characteristic anisotropy in Fourier space. Hence 
the difference between $\hat{E}$ and $\hat{B}$ can be
taken as a measure of the cosmological signal as compared to the
foregrounds and/or instrumental offsets. If foregrounds and not noise
are expected to be the main limitation then one may use
may subtract the power spectrum
of $\hat{E}$ from the one for $\hat{B}$
in equation \ref{optimal} 
and no further foreground removal is
needed. One could therefore use this technique to measure
polarization even at frequencies below 50GHz, where foreground contribution
is large, but could be averaged over if sufficient number of channels
are being measured. 
Note that this test does not depend on 
the temperature anisotropies at all and can be applied directly 
to the measurements of $Q$ and $U$. To obtain a statistically significant
measure of polarization one only needs to show that in an rms sense
$\hat{E}$ is larger than $\hat{B}$.
If one is noise and not foreground limited then it is more advantageous
to use the optimal estimator in equation \ref{optimal}, 
because subtracting the power spectrum of 
$\hat{B}$ from the optimal estimator for $\hat{E}$ leads to an increase
in noise by $2^{1/2}$.
In practice the actual analysis will depend on the
level of foregrounds and other systematics (e.g. sidelobe pickup) 
relative to noise and different 
estimators must be tested for consistency, but it is 
important to note that in the case of polarization 
we have a possibility to use a combination of Stokes parameters in which
foregrounds can be separated from the cosmological signal, something 
that cannot be achieved in the temperature measurements alone.

\section{Tensor polarization}

Discussion in \S 4 applies only if polarization is produced by scalar 
perturbations. While this is certainly a valid approximation on 
small angular scales ($l>100$), on larger scales one may be
able to detect polarization from non-scalar perturbations,
induced either by vector or tensor modes. 
The latter are of particular interest, because they are expected to be
present in several inflationary based models, although only on
large angular scales and with rather small amplitudes (Crittenden 
et al. 1993). Defect models also predict production of both tensor
and vector perturbations.
In the case of tensor
perturbations temperature anisotropy and the two Stokes parameters
can be written in the small scale limit as (Bond 1996, Kosowsky 1996),
\begin{eqnarray}
T^{(T)}(\vec{n})&=&\int d^3\vec{k}\left(1-\mu^2\right)\Delta_T^{(T+)}(k,\mu)
\nonumber \\
Q^{(T)}(\vec{n})&=&\int d^3\vec{k}\left[\left(1+\mu^2\right)\Delta_P^{(T+)}(k,\mu)
\cos(2\phi_{\vec{k}})
-2\mu
\Delta_P^{(T\times)}(k,\mu)\sin(2\phi_{\vec{k}})\right]
\nonumber \\
U^{(T)}(\vec{n})&=&\int d^3\vec{k}\left[\left(1+\mu^2\right)\Delta_P^{(T+)}(k,\mu)
\sin(2\phi_{\vec{k}})
+2\mu
\Delta_P^{(T\times)}(k,\mu)\cos(2\phi_{\vec{k}})\right]
,
\label{tens}
\end{eqnarray}
where $\Delta_P^{(T+)}(k,
\mu)$ and $\Delta_P^{(T\times)}(k,\mu)$ are the two independent 
polarization states of a gravity wave with equal rms amplitudes.
For convenience we defined
the orientation of the two polarization states 
in the plane perpendicular to $\hat{k}$ 
so that local $x$ direction points in the direction of the pole, in which case
only $\Delta_T^{(T+)}(k,\mu)$ contributes to the temperature in the small scale limit. 
Expectation values for $\Delta_P^{(T+)}(k,\mu)$ and $\Delta_P^{(T\times)}(k,\mu)$
can be calculated just like in the scalar case by expanding them into 
a Legendre series and solving 
a system of Boltzmann equations
(Crittenden et al. 1993) or the integral solution  
(Zaldarriaga \& Seljak 1996). Because of additional $\mu$ terms in equation 
\ref{tens} a more
convenient set of variables is obtained by eliminating the explicit 
$\mu$ dependence (Kosowsky 1996),
\begin{eqnarray}
B_l^{1,\epsilon}&=&{2 \over 2l+1}\left[(l+1)\Delta_{P,l+1}^{(T\epsilon)}+l\Delta_{P,l-1}^{(T\epsilon)}
\right]
\nonumber \\
B_l^{2,\epsilon}&=&
{1 \over 2l+1}\left[{(l+1)(l+2) \over 2l+3}\Delta_{P,l+2}^{(T\epsilon)}
+2{6l^3+9l^2-l-2 \over (2l-1)(2l+3)}\Delta_{P,l}^{(T\epsilon)}
+{(l-1)l \over 2l-1}\Delta_{P,l-2}^{(T\epsilon)}\right],
\end{eqnarray}
where $\epsilon$ stands for $+$ and $\times$ and all the 
variables explicitly depend on $k$.

One can now follow the same steps 
as in the case of scalar perturbations, which transform the angle
$\phi_{\vec{k}}$ into $\phi_{\vec{l}}$. 
The variables $E^{(T)}$ and $B^{(T)}$ defined in equation \ref{fore} 
(where superscript $T$ indicates that these are produced by
tensor
modes) are independent of the azimuthal angle and the two 
tensor components decouple, so that $\Delta_P^{(T+)}(k,
\mu)$ contributes only to $E^{(T)}$ and $\Delta_P^{(T\times)}(k,\mu)$
contributes only to $B^{(T)}$. Their power spectra
can be expressed in terms of integrals over $B_l^1$, $B_l^2$ as,
\begin{eqnarray}
\langle B^{(T)}(\vec{l}) B^{(T)*}(\vec{l'}) \rangle &=&
(2 \pi)^2\delta(\vec{l}-\vec{l}')\int d^3k |B_l^1|^2(k) \nonumber \\
\langle E^{(T)}(\vec{l}) E^{(T)*}(\vec{l'}) \rangle &=&
(2 \pi)^2\delta(\vec{l}-\vec{l}')\int d^3k |B_l^2|^2(k) \nonumber \\
\langle E^{(T)}(\vec{l}) B^{(T)*}(\vec{l'}) \rangle &=&0.
\end{eqnarray}
The cross correlation term vanishes 
because the two tensor polarization states are
independent.
The variable $B^{(T)}$ does not vanish in the case of tensor 
perturbations and its power spectrum 
differs from the power spectrum of $E^{(T)}$. 
Note that 
different combinations of $Q$ and $U$ will result in different power 
spectra,
which can always be expressed in terms of the two defined 
above. Just like the cross term between $E^{(T)}$ and $B^{(T)}$
vanishes so does the cross correlation term between $T^{(T)}$
and $B^{(T)}$.
There is only one power spectrum present in the
case of temperature-polarization cross correlation, 
\begin{eqnarray}
\langle T^{(T)}(\vec{l}) B^{(T)*}(\vec{l'}) \rangle &=&
0\nonumber \\
\langle T^{(T)}(\vec{l}) E^{(T)*}(\vec{l'}) \rangle &=&
(2 \pi)^2\delta(\vec{l}-\vec{l}')\int d^3k \Delta^{(T)}_{Tl}(k)B_l^2(k).
\end{eqnarray}
Detailed calculations of these spectra have been presented elsewhere 
(Seljak \& Zaldarriaga 1996b).

\section{Model predictions}

\begin{figure}[t]
\vspace*{17.3 cm}
\caption{ Power spectra of 
polarization (a), temperature-polarization cross correlation
(b), temperature (c) and correlation coefficient (d). The models
are standard CDM (solid curve), open CDM (dotted curve) and 
reionized standard CDM with optical depth of 0.2 (dashed curve).}
\includegraphics{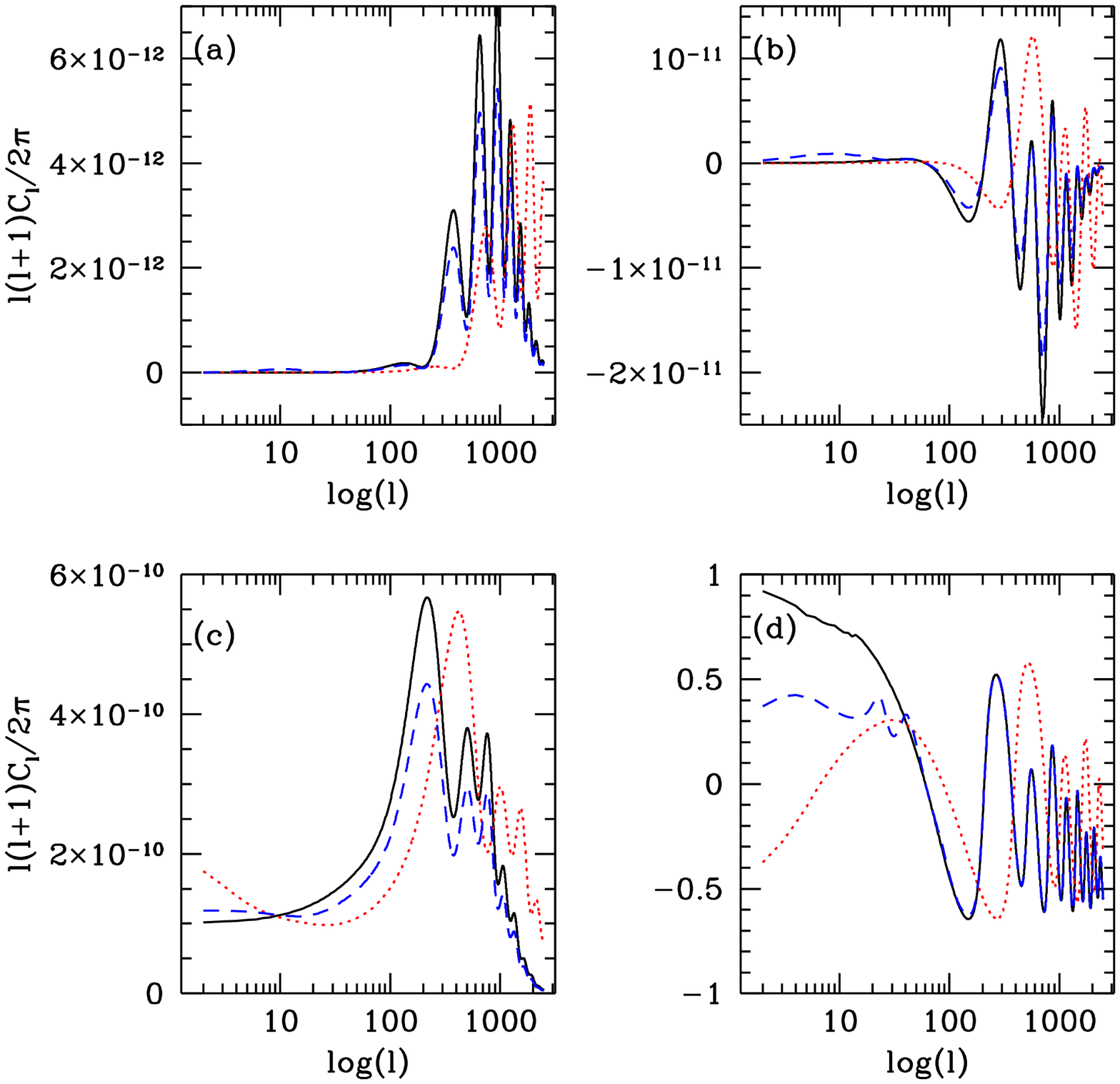}
\label{fig1}
\end{figure}

Instead of the power spectrum $C_l$ we will use the quantity 
$l(l+1)C_{l}/2\pi$, 
which gives the contribution to the variance
per logarithmic interval of $l$. This is a familiar quantity in the 
case of temperature anisotropies, where its broad band average is
approximately flat up to 
the damping scale. 
Predictions for $l(l+1)C_{Pl}/2\pi$ and $l(l+1)C_{Cl}/2\pi$ are given in 
figures \ref{fig1}a,b
for a variety of cosmological models. For comparison we also plot the
usual $l(l+1)C_{Tl}/2\pi$ in figure \ref{fig1}c, as well as the correlation 
coefficient in figure \ref{fig1}d.
All the 
model predictions have been computed using CMBFAST package by Seljak \&
Zaldarriaga (1996). One can see that typically the models predict very
little polarization on large angular scales, below $l \sim 200$. 
On smaller angular scales most of the models predict
polarization at the level of 5-10$\mu$K. There are several characteristic
features of interest in this regime. The most important one is that the
acoustic peaks are narrower than the corresponding ones in temperature
anisotropy. One can understand this with the help of tight coupling
approximation (Hu \& Sugiyama 1995; Seljak 1994; Zaldarriaga \& 
Harari 1995). The dominant source of temperature anisotropy are
intrinsic
photon anisotropy ($\Delta_0$) and velocity ($\Delta_1$). 
Both terms oscillate, but
are out of phase with each other. This means that they partially cancel
each other and oscillations in the temperature anisotropy
are less pronounced than they would be if
only one term were contributing. The dominant source of polarization
is photon second moment $\Delta_2$, so the
oscillations are more pronounced than in the case of temperature
anisotropy. These oscillations are even more pronounced in the case
of temperature-polarization cross-correlation, which can be either 
positive or negative. 
Another characteristic of polarization is that it is not sensitive to
the integrated Sachs-Wolfe term. This term is responsible for increase
in the temperature anisotropies at low $l$, as in models with 
cosmological constant, curvature or if recombination occurs
close to the matter-radiation equality, during which gravitational 
potential is changing with time. Another class of such models are 
topological defect models, where small $l$ spectrum is dominated by 
late time integrated Sachs-Wolfe effect. A measurement of polarization 
at a few microkelvin level will put a significant constraint on such models.

One of the parameters that are of special importance for polarization
is the optical depth to Thomson scattering. As photons propagate 
through the universe they scatter off free
electrons, which were ionized by UV light either from an early generation
of stars or from quasars. Current limits give that the universe was 
mostly ionized up to the redshift of 5, which results in optical depth of 
the order of 1\% in standard CDM universe and somewhat larger in open or
high baryon models. 
It is likely that the reionization did not 
occur much earlier so that the optical depth would exceed unity, because 
then it would suppress CMB anisotropies on small angular scales, 
in contrast with recent observational data (Netterfield et al. 1995, Scott
et al. 1996). 
The exact 
epoch of reionization remains however unknown and its determination
would provide an important constraint to the models of galaxy 
formation. Temperature anisotropies alone 
will not be able to constrain this epoch significantly, because even in 
the most optimistic scenario sensitivity to optical depth $\tau$ is 
around 10-20\% (Jungman et al. 1996). This is because reionization
is degenerate
with the amplitude of fluctuations, 
which can be only broken at low $l$, where 
cosmic variance is large. 
Polarization can
help here both because reionization introduces new structure and also 
simply because the cosmic variance is reduced as more independent
realizations are observed.
As shown in figure \ref{fig1}, the effect of reionization is to
increase somewhat the amplitude of polarization at low $l$, but not by
much and the amplitude still remains below 1-2$\mu K$. The effect is better
seen in the cross-correlation spectrum and in the corresponding 
correlation coefficient. The latter clearly 
displays the rich structure
at low l that allows one to determine epoch of reionization and the 
integrated optical depth (Zaldarriaga 1996). 
On smaller angular scales polarization
amplitude decreases with optical depth just like the temperature 
anisotropy and the ratio of the two remains constant (Bond \& Efstathiou 
1987). 

\begin{figure}[t]
\vspace*{17.3 cm}
\caption{Variance in polarization (a) and cross-correlation 
power spectrum (b) for a satellite with noise characteristic 
$w_P=w_T/2$. We assumed 50\% sky coverage and $0.3^\circ$ degree
beam ($0.2^\circ$ in the most optimistic case). 
The spectra were averaged over a 20\% band in $l$ and the bands shown 
are one standard deviation above and below the  
underlying model, taken to be
COBE normalized standard CDM. }
\includegraphics{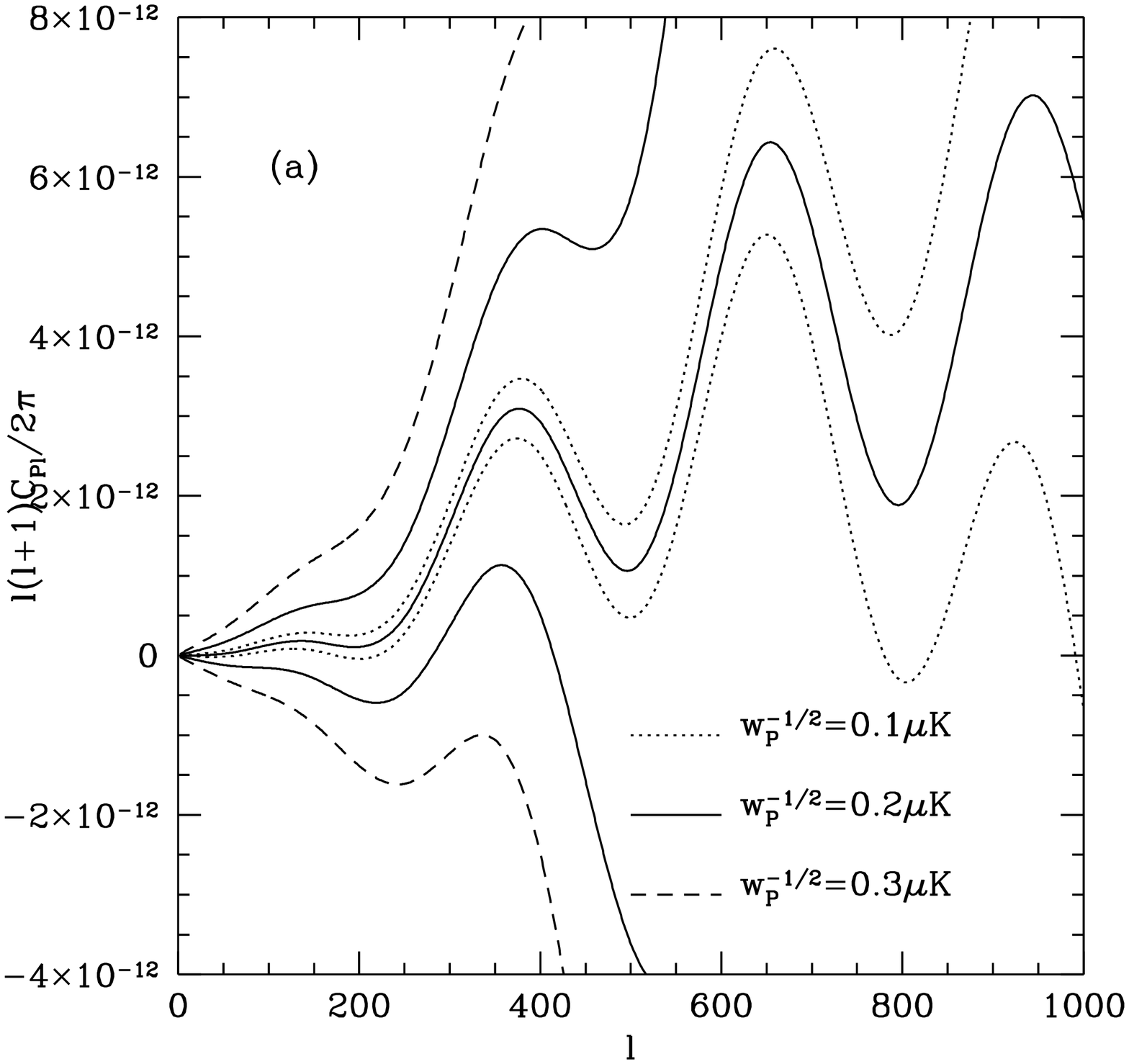}
\includegraphics{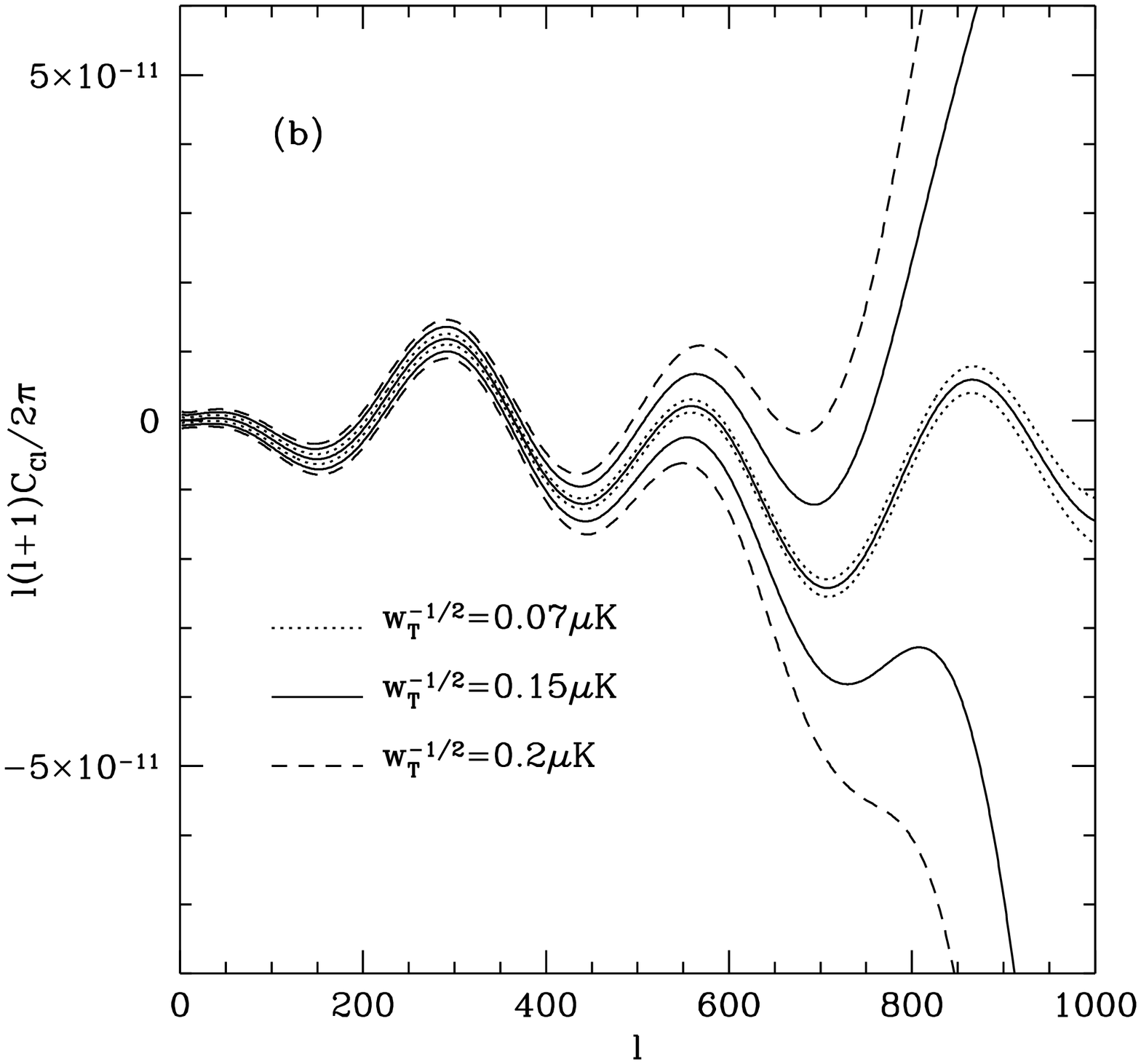}

\label{fig2}
\end{figure}

Figure \ref{fig2} presents a more quantitative estimate of sensitivity 
in the case of satellites, assuming standard CDM as the underlying model. 
The middle curve shown is the underlying theoretical model while the 
two curves above and below show the one standard deviation of the 
reconstructed spectrum from the true model.
The variances were obtained using equations \ref{covpl} and \ref{covcl},
assuming 50\% sky coverage ($\Omega=2\pi$) and
$d\ln l=0.2$. We adopted 3 different noise characteristics. The most 
optimistic possibility is $w_P^{-1/2}=0.1\;\mu$K in $0.2^{\circ}$
beam, which could easily be achieved by COBRAS/SAMBA satellite with 
their bolometric detectors. The intermediate model 
assumes $w_P^{-1/2}=0.2\;\mu$K and $0.3^{\circ}$ beam, which could be
feasible with the MAP satellite by combining their most sensitive channels.
The third model is the most conservative one and assumes 
$w_P^{-1/2}=0.3\;\mu$K and $0.3^{\circ}$. All of the sensitivities 
assume 1 year of observation and longer observation periods would 
reduce the noise accordingly. From figure \ref{fig2} we see 
that only the most optimistic model is capable of constraining the 
polarization power spectrum significantly. On the other hand, for
cross-correlation spectrum the situation is much better and all of the
assumed models will give some positive detection, although of course
with lower noise levels one will be able to extend this to much 
smaller angular scales. This difference between polarization and
cross-correlation is to be expected, because noise in 
temperature is lower and because cross correlation power spectrum 
has more power than the polarization power spectrum itself. 
Although more detailed calculations are needed to estimate the
sensitivity for actual satellites (Zaldarriaga et al. 1996), 
it is clear that reducing the
noise by a factor of 2 leads to a significant improvement in the 
sensitivity to polarization. 

\section{Conclusions}

Polarization in cosmic microwave background has the promise to become
the new testing ground of processes in the early universe quite 
independent of the measurements of temperature anisotropies. The 
spectrum of polarization, while lower in signal than the temperature
anisotropies, can be more sensitive to certain parameters such as 
reionization or gravity waves. 
Even for the determination
of the standard parameters polarization provides some advantages,
for example, acoustic oscillation peaks are much more prominent 
and so can be more easily detected. In addition to the spectrum of
polarization one can also determine the spectrum of temperature-polarization
cross-correlation. Two additional spectra  
can help to break some of the degeneracies present in the estimation 
of cosmological parameters. This is specially important 
for those parameters whose precision is limited by cosmic 
variance, because polarization provides additional independent
realizations of initial conditions in the universe.

Even though measurements in polarization are not likely to 
achieve the same level of precision as in the case of temperature
anisotropies,
polarization still has some advantages that may prove crucial if
the amplitude and complexity of galactic foregrounds and extragalactic
point sources have been significantly underestimated. 
Multifrequency subtraction is simpler in the case of polarization,
because fewer foregrounds are polarized and need to be
modeled. 
More importantly, polarization induced by scalar perturbations
has a unique signature in Fourier space and by exploiting this one may
separate cosmological signal from other sources of polarization. 
Although the analysis in this paper has been limited to small scales it
has recently been shown that the same property is also valid in the more 
general all-sky analysis (Zaldarriaga \& Seljak 1996; Kamionkowski, 
Kosowsky \& Stebbins 1996). This signature
would be especially important if the level of foregrounds is 
significantly larger than expected. While one would not be 
able to remove the foregrounds from the temperature maps, a combination
of $Q$ and $U$ Stokes parameters would allow one to subtract statistically 
the effects
of foregrounds the case of polarization.
This technique would be feasible
both for interferometer measurements or for 
measurements with a large coverage of the sky such as the forthcoming
satellite missions. 
The sensitivity of satellites will be sufficient for 
an unambiguous determination of polarization and indeed polarization may
prove to be crucial to break some of the degeneracies present in the
parameter reconstruction from the temperature anisotropies alone 
(Zaldarriaga et al. 1996). 
Finally, if foregrounds can be controlled, then a unique signature of
tensor (or vector) perturbations could be directly observed, although
this would require an exquisite understanding of 
noise properties, systematics and foregrounds 
at the level of 0.5$\mu K$. 
Given
the unique nature of information present in the microwave background 
and its simple linear 
depence on the underlying cosmological parameters
it is important to explore it 
at its maximum, which certainly includes polarization as one of its
main components.

\acknowledgements
I would like to thank Douglas Scott, 
Arthur Kosowsky, David Spergel, Martin White and specially Matias Zaldarriaga 
for useful discussions. 
\appendix
\section{Appendix}

In this Appendix we show how the small scale limit of correlation
functions $QQ$ and $UU$ (equations \ref{corrf}) follows from the expression
derived by Coulson et al. 1994. We start from their expression
(correcting for missing factors of 1/2),
\begin{eqnarray}
\langle Q(0)Q(\vec{\theta})\rangle &=&\sum_l {2l+1 \over 8 \pi}\left[
C_{Pl}P_l(\cos\theta)+{\cos(4\phi) \over 2}
C_{PPl}P_l^4(\cos \theta)\right] \nonumber \\
C_{PPl}&=&{(l-4)! \over (l+4)! } \sum_{l'}(2l'+1)\int k^2dk \Delta_{Pl'}
\Delta_{Pl}a^4_{l'l},
\end{eqnarray}
where $P_l$ is the Legendre polynomial and $P_l^4$ the associated Legendre
function of 4th order.
Coefficients $a^4_{ll'}=\int_{-1}^1dxP_l(x)P_{l'}^4(x)$ have a closed form
expression (Coulson 1994) and in the large $l$ limit they 
peak
at $l=l'$. More importantly, $\Delta_l$ is a rapidly oscillating function of
$k$
and for $l\ne l'$ the integral over $k$ leads to almost complete
cancellation of $\int k^2dk \Delta_{Pl'}
\Delta_{Pl}$. 
Thus in this limit
one can write
\begin{equation}
C_{PPl}={2l(l-1)(l-2)(l-3)(l-4)! \over (l+4)! }C_{Pl}, 
\end{equation}
where we used 
the $l=l'$ closed form for of $a^4_{ll'}$ (Coulson 1994).
Furthermore, in the $l \rightarrow \infty$ limit Legendre functions
can be written as $P_l^m(\cos\theta)=J_m(l\theta)l^{-m}(l+m)!/(l-m)!$,
where $J_m(x)$ is the Bessel function of order $m$. Combining all the
expressions leads to $\langle Q(0)Q(\vec{\theta})\rangle$ correlation
function given in
equation \ref{corrf}. Other correlation functions in equation \ref{corrf}
can be derived from
the expressions in Coulson et al. 1994 using similar manipulations.
Numerical results presented in 
Zaldarriaga \& Seljak 1996 show that small  
scale approximation derived here is an excellent approximation everywhere 
except at very 
small $l$ ($l\lsim 20$).


\end{document}